\documentclass[a4paper,11pt]{article}
\usepackage{jinstpub} 
\usepackage{subfig}
\usepackage{tablefootnote}

\title{\boldmath A search for a muon to electron conversion in COMET}

\author{Yuki Fujii}
\collaboration[c]{on behalf of the COMET collaboration}
\affiliation{Monash University,\\
Clayton, Victoria, Australia}

\emailAdd{yuki.fujii@monash.edu}

\abstract{The COMET experiment at J-PARC, Japan, aims to search for muon to electron conversion with aluminium nuclei, achieving a sensitivity four orders of magnitude higher than the current upper limit at a 90\% confidence level.
The proton beam line has recently been completed, and muons have been successfully transported through the curved solenoid in the Phase-alpha of the experiment.
In this paper, we will present preliminary results from the Phase-alpha beam measurement, the status of the intermediate sensitivity experiment (COMET Phase-I), and the ultimate goal of COMET Phase-II.}


\newcommand{\muec}{\mu^- N \to e^- N}
\newcommand{\muecS}{\mu\text{--}e}

\begin{document}
\maketitle
\flushbottom

\section{Introduction}
\label{sec:intro}

The COMET experiment searches for a transition of a muon to an electron without emission of neutrinos, also known as a $\muecS$ conversion.
This process violates the conservation law of lepton flavour and is strictly forbidden in the standard model of particle physics (SM).
Even when considering the minimum SM model extension with neutrino oscillations, the conversion rate ($CR(\muec)$) is strongly suppressed to approximately $10^{-54}$, which is completely undetectable.
On the other hand, a large enhancement of the $CR(\muec)$, up to $10^{-15}$, is predicted in many BSM models, e.g., Leptoquark models \cite{leptoquark}.
The signal electron is essentially monochromatic with an energy $E_{\mu e} = M_\mu - B_\mu - E_{recoil} \approx$ 105~MeV in case of muonic aluminium.
This peculiar feature together with the low background level make the $\muecS$ conversion an ideal tool to investigate the BSM physics.
The current upper limit on $CR(\muec)$ is $7\times10^{-13}$ at 90\% confidence level set by the SINDRUM II experiment \cite{sindrum2}.

In COMET, our primary target is to achieve 100 times better single event sensitivity, $10^{-15}$, which is called COMET Phase-I \cite{tdr}, followed by another two orders of magnitude improved sensitivity in Phase-II.
This allows us to indirectly explore the new physics energy scale up to 100--1000~TeV.
Furthermore, the complementary searches with other muon CLFV modes ($\mu \to e\gamma$ and $\mu \to eee$) will provide more detailed new physics investigation by looking at slightly different parameter space.
In addition to those two stages, we recently performed the first muon beam delivery to the COMET experimental area with a set of detectors, which is called COMET Phase-alpha.
In this paper, we first introduce the event feature of the signal and backgrounds, then we report details of the experimental apparatus used in Phase-I and Phase-II, and the current preparation status.
Also, the preliminary results from the first muon beam delivery and profile measurement in COMET Phase-alpha, are briefly presented.

\section{Signal and Backgrounds in COMET}
In the SM, a muon decays into an electron and two neutrinos to conserve the lepton flavour; $\mu^- \to e^- \nu_\mu \overline{\nu_e}$.
If this happens while the muon is bounded to an atom, this is known as muon decay in orbit (DIO).
Due to the presence of a nucleus, DIO electrons receive the recoil energy and they can have a broad energy spectrum with an endpoint of $\approx$105~MeV, hence, the DIO is an intrinsic background for the signal event.
Alternatively, a muon is captured by a nucleus and emits a neutrino and neutrons/protons/alpha particles.
The capture process also includes a radiative process called radiative muon capture; RMC.
This process produces a photon or a pair of $e^+e^-$ in the final state.
However, the maximum energy of gamma-rays is sufficiently lower than the of the conversion electron energy to make this background negligible.
The DIO events can only be discriminated by their energy, therefore the precise momentum measurement is a key design factor.
According to our studies based on simulations, a momentum resolution of 200~keV/$c$ is required to achieve a single event sensitivity of $10^{-15}$.
To lower the sensitivity by a factor of 100, this resolution has to be improved further down to 150~keV/$c$.
In the case of the signal event, a muon stops inside the target and forms a muonic atom at $1s$ ground state.
The electron emitted from the muonic aluminium has a monochromatic energy of 105~MeV as already explained with a decay time of $\tau_\mu =$ 864~ns.

Other sources of background are the electrons produced by the in flight decays of muons or pions.
In this case, the time delay due to the muonic atom lifetime is not present so this prompt background can be suppressed by a timing cut.
We therefore adopt a pulsed beam structure with a separation of $\sim$1.2~$\mu$s between the pulses and use a delayed measurement window between 700~ns and 1170~ns from the bunch arrival on the target to avoid those prompt backgrounds in COMET.
In this scheme, it is also important to maintain a highly pulsed structure, since any leakage protons appearing inter-bunches can create a signal-like secondary particle in this time window.
We call this an ``Extinction'' factor defined as; $R_\mathrm{ext}=\mathrm{(\#of \ protons \ between \ bunches)}/\mathrm{(\#of \ protons \ in \ a \ main \ bunch)}$, and this $R_\mathrm{ext}$ should be less than $10^{-10}$ ($10^{-11}$) to accomplish the Phase-I (Phase-II) sensitivity.

The last background is a cosmic-ray induced background randomly produced inside the detector with an energy close to 105 MeV.
The cosmic-ray background will be suppressed by covering the entire detector with a cosmic-ray-veto detector.

\section{Apparatus}\label{sec:apparatus}
\begin{figure}[htbp]
\centering
\subfloat[\centering Conceptual view of the staging approach]{{\includegraphics[width=.53\textwidth]{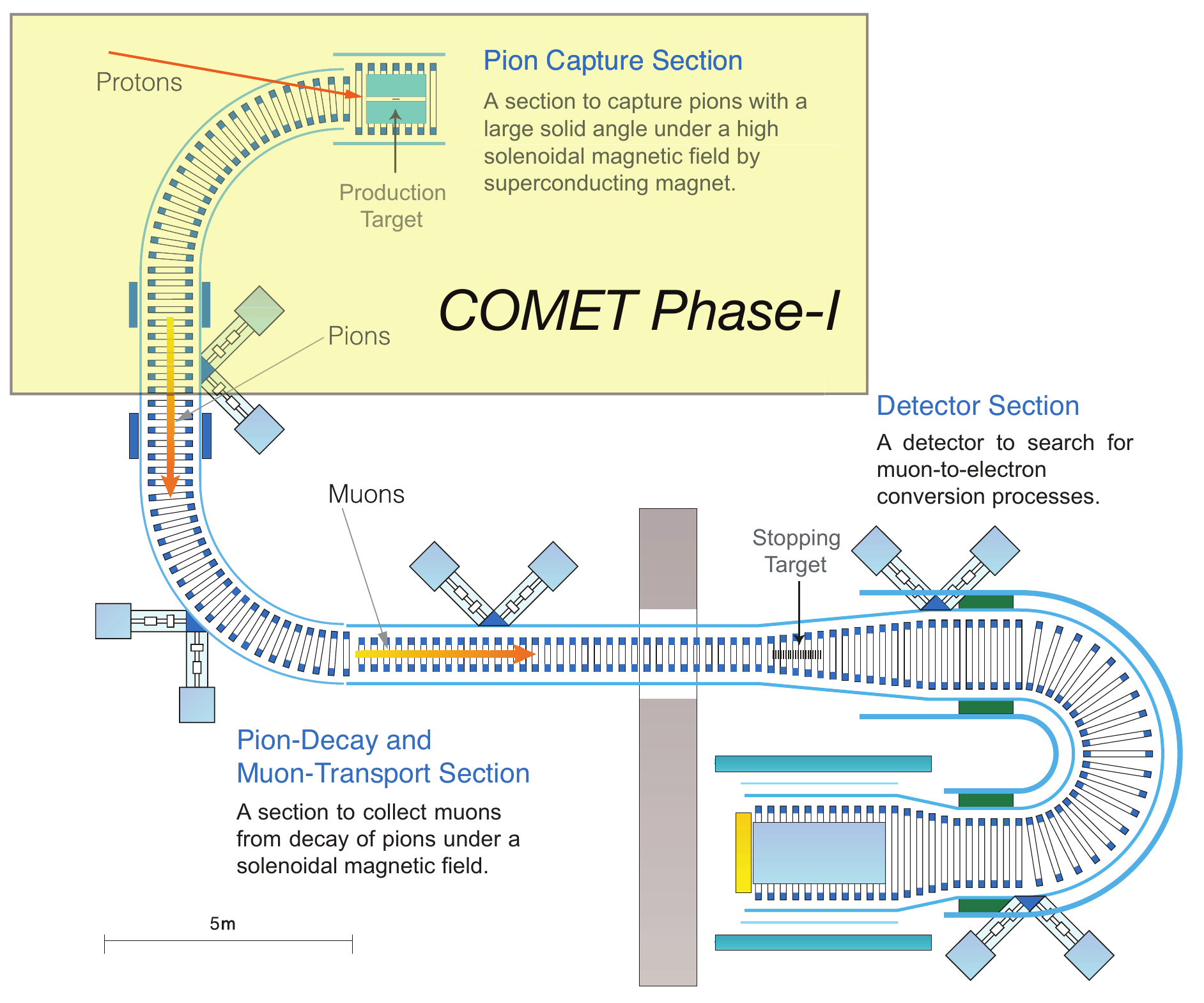}}}
\qquad
\subfloat[\centering Detailed Phase-I layout]{{\includegraphics[width=.32\textwidth]{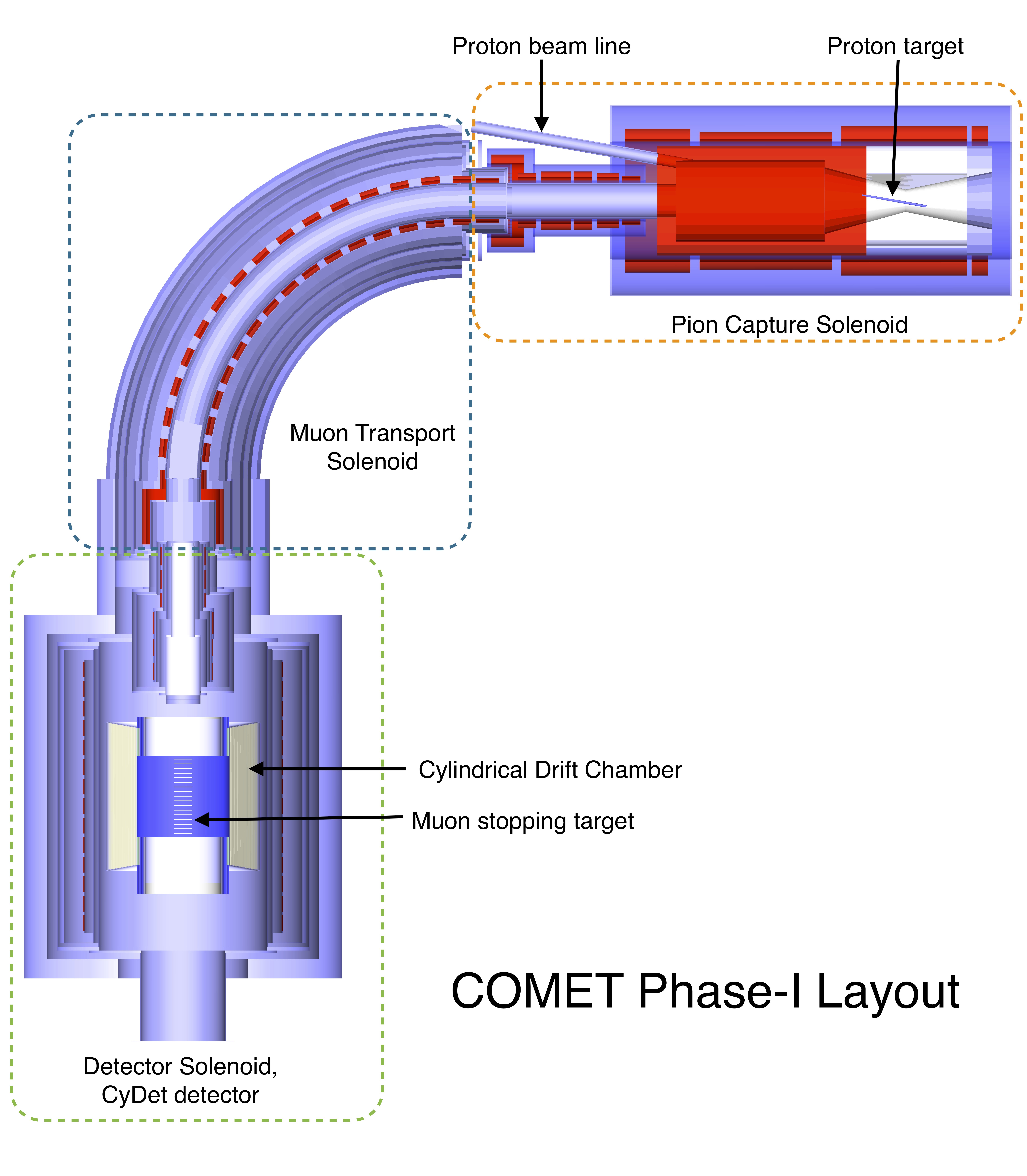}}}
\caption{Overview of the COMET Phase-I and Phase-II conceptual design.\label{fig:comet_concept}}
\end{figure}
\paragraph{Muon beam}
To produce the world's most intense muon beam, we use the proton beam accelerated up to 8~GeV in the main ring (MR) of the proton synchrotron at J-PARC (Japan Proton Accelerator Research Complex), Japan.
To realise an ideal bunch separation timing, only one of two buckets is filled with protons at the rapid cycling synchrotron which accelerates the proton beam from 400~MeV to 3~GeV before the MR.
This corresponds to 4 out of 9 total buckets in the main ring with a bunch separation of 1.17~$\mu$s.
After the acceleration, protons are slowly extracted towards the J-PARC Hadron experimental facility by using one or two electrostatic septa (ESS) and multiple magnetic septa (SMS) while keeping the bunched structure (BSX: Bunched Slow Extraction) \cite{SX}.
The protons are delivered to the COMET beam area and impinged into the pion production target which is surrounded by the 5~T pion capture solenoid.
Through the hadronic interaction, pions are produced, and low momentum back scattered pions are efficiently transferred towards the 3~T muon transportation solenoid (MTS).
Most of pions decay into muons before reaching the exit of the MTS.
Due to the curved structure of the MTS, charged particles vertically drift depending on their charge and transverse momentum while travelling inside,
with an additional dipole field used to adjust compensate the drifting of low momentum negative muons to keep them around the centre at the end of the MTS.

\begin{figure}[htbp]
\centering
\subfloat[\centering CyDet]{{\includegraphics[width=.35\textwidth]{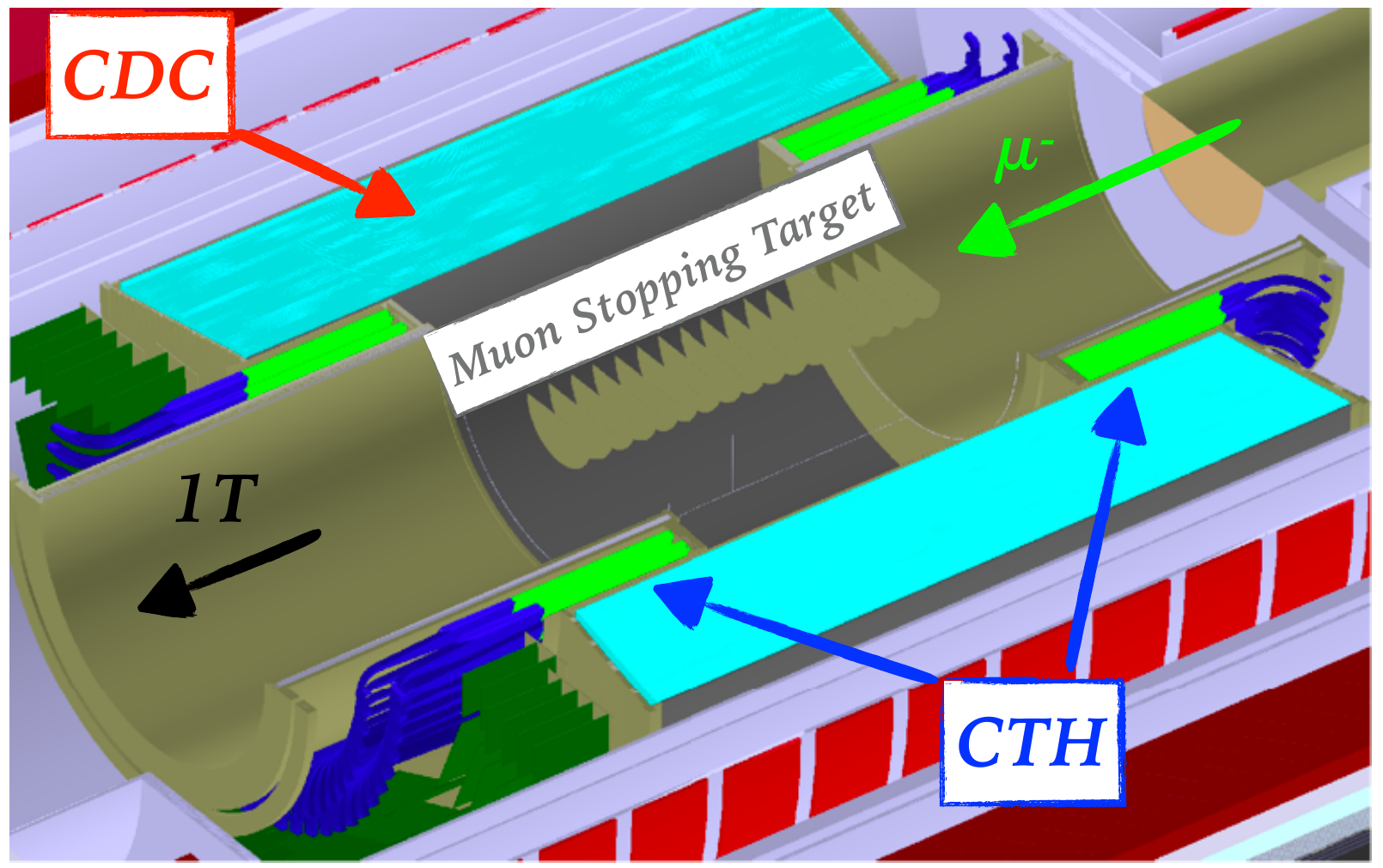}}}
\qquad
\subfloat[\centering StrECAL]{{\includegraphics[width=.5\textwidth]{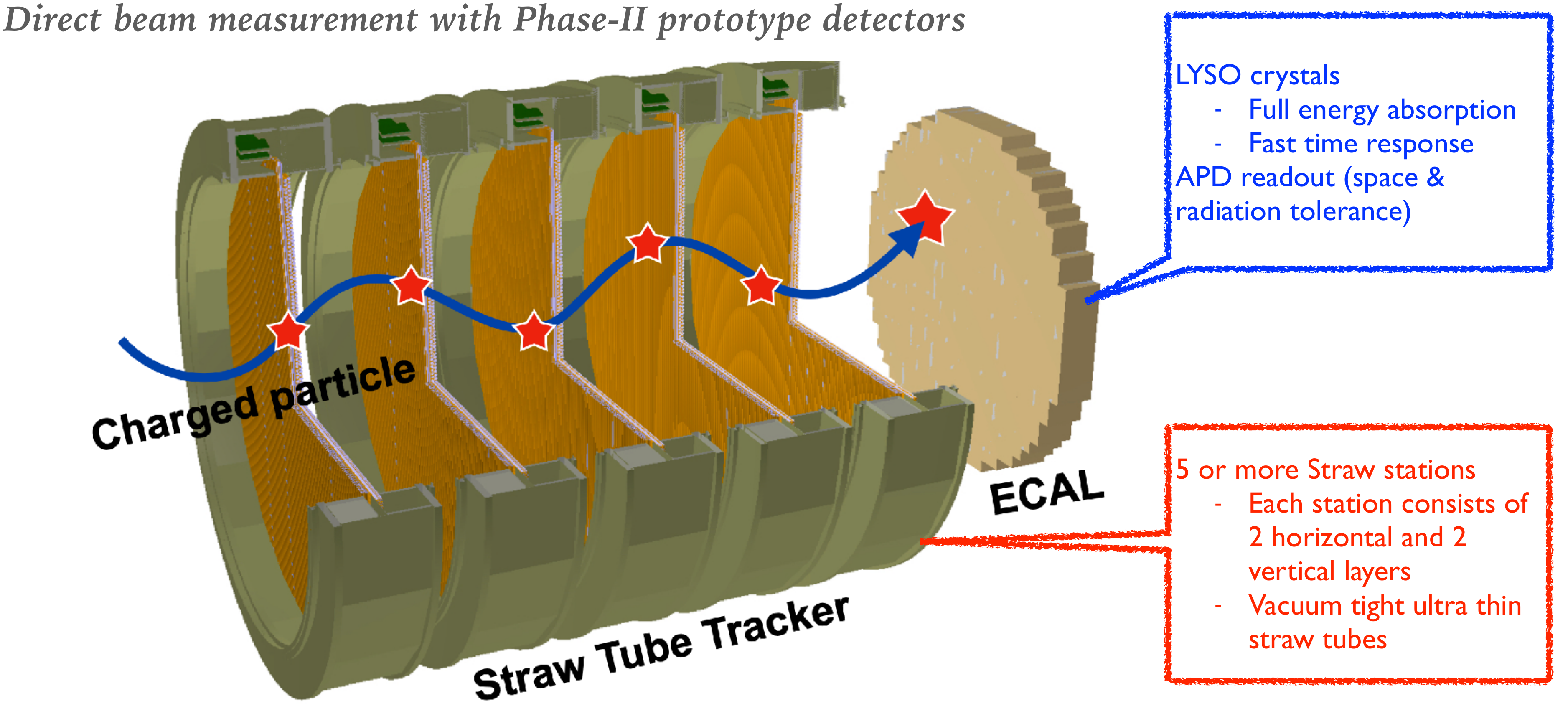}}}
\caption{Two detector systems used in COMET Phase-I.\label{fig:detectors}}
\end{figure}

\paragraph{Phase-I}
In Phase-I, we will conduct the physics measurement using a Cylindrical Detector (CyDet) system placed at the end of the first 90 degrees of the MTS (see Figure~\ref{fig:comet_concept}) inside the 1~T solenoidal magnetic field (Detector Solenoid: DS).
The muon stopping target, a series of 200~$\mu$m thick aluminium disks, is located inside the CyDet.
The cylindrical shape helps to avoid the remaining beam particles such as pions and their secondaries that concentrate on the beam axis with small transverse momentum, as well as filtering the most of DIO electrons whose momentum is below 70~MeV/$c$.
The CyDet system consists of a Cylindrical Drift Chamber (CDC) and a set of Cylindrical Trigger Hodoscope (CTH) detectors as shown in Figure~\ref{fig:detectors}(a).
The CDC, filled with a He:$i$C$_5$H$_{10}$ (90:10) gas mixture, contains almost 5,000 all stereo anode wires, and is used to reconstruct the momentum of charged particles \cite{cdc}.
The wires are read out by 105 RECBE boards \footnote{Readout Electronics for CDC in BElle (RECBE) \cite{recbe}} with minor adaptations.
The expected momentum resolution is about 200~keV/$c$, which is good enough to achieve the Phase-I physics sensitivity.
The CTH detectors are located at both ends of CDC and consist of two layers of concentric rings containing 64 counters each \cite{cth}.
Each counter is a plastic scintillator (BC-408) with height$\times$width$\times$length of 5 $\times$ 88 $\times$ 360 (10 $\times$ 88 $\times$ 340)~mm$^3$ for the inner (outer) layer.
All scintillators are connected to fibre bundles to readout the scintillation photons outside of the detector solenoid for using silicon photo-multipliers (SiPM) which are relatively weak against radiation damage and have to be placed outside of the detector solenoid where the neutron level is an order lower, inside the cooling box.
Due to an extremely high hit rate environment, the fake trigger could be an issue especially in Phase-I physics measurement.
To suppress the trigger rate caused by non signal-like fake events as low as possible,
an online trigger system is introduced by utilising machine learning based algorithms inside field programmable gate arrays (FPGAs) in the CyDet system, which is called COTTRI (COmeT TRIgger) \cite{cottri}.
This will realise the trigger rate of less than 13~kHz, which is within the maximum available rate for our data acquisition (DAQ) system.

We will also perform a direct beam measurement by replacing the CyDet with a set of planar shaped detectors called StrECAL (Straw+ECAL, see Figure~\ref{fig:detectors}(b)).
The beam measurement detector consists of a series of straw tube tracker (StrawTrk) stations and a full absorption type electron calorimeter (ECAL).
The StrawTrk has five stations and each one consists of one vertical layer and the other horizontal layer.
Each layer has staggered straw tubes in order to minimise gaps to reduce inefficiency.
A straw is made of 20~$\mu$m thick aluminised Mylar with a 10~mm diameter, filled with Ar:CO$_2$=50:50.
The ECAL is composed of arrays of LYSO (Lu$_{2(1-x)}$Y$_{2x}$SiO$_5$) crystals with dimensions $2\times2\times12$~cm$^3$ \cite{ecal}.
This length corresponds to 10.5 radiation lengths to contain the signal like electrons with minimum leakage.
In each crystal, scintillation photons will be read by an avalanche photodiode (APD, Hamamatsu S8664-1010, $10\times10$~mm$^2$).
Both StrawTrk and ECAL signals are read by specially designed waveform digitisers \cite{strecalRO} based on domino ring sampler chips (DRS4).
The beam measurement detectors are also regarded as detector prototypes for Phase-II physics measurement.
The DS will be covered by the iron yoke and cosmic ray veto counters made of plastic scintillator.
Also, there is a possibility to introduce resistive plate chambers (RPC) around the MTS to DS bridging solenoid where plastic scintillators cannot work properly due to higher neutron levels expected.

\paragraph{Phase-II}
To achieve another 100 times better sensitivity than in Phase-I, the production target will be replaced from graphite to tungsten to increase the pion yield, and shielding materials will be reinforced.
The MTS will be extended by another 90 degrees to further reduce pions, and there will be an additional electron spectrometer after the muon stopping target to filter most DIO electrons.
This electron spectrometer enables using the detector system that covers almost fully 2$\pi$ forward direction, increasing the signal acceptance.
In COMET Phase-II, it is crucial to improve the momentum resolution which is dominated by the multiple scattering effects.
Therefore, all straw tubes will be replaced by thinner ones,  12~$\mu$m wall thickness with twice smaller diameter of 5~mm to reduce the material budget inside the tracking volume.
The ECAL will cover the 50~cm radius to increase the acceptance with 1,920 crystals.

\section{Physics Sensitivity}
In Table~\ref{tab:phys}, we summarise the breakdown of the expected signal acceptance in each phase together with the expected number of background events.
\begin{table}[htbp]
\centering
\caption{Signal efficiency and expected number of background events for the COMET Phase-I \cite{tdr} / Phase-II \cite{kou_th,ben_th}.\label{tab:phys}}
\smallskip
\begin{tabular}{l|c|c}
\hline
Items & Phase-I & Phase-II \\
\hline
Signal acceptance & 0.2 & 0.18 \\
Trigger\,+\,DAQ efficiency & 0.8 & 0.87 \\
Track finding efficiency & 0.99 & 0.77 \\
Track selection & 0.9 & 0.94 \\
Momentum window & 0.93 & 0.62 \\
Timing window & 0.3 & 0.49 \\
\hline
Total & 0.04 & 0.034 \\
\hline
Sources of background &  &  \\
\hline
Muon decay in orbit & 0.01 & 0.068 \\
Radiative muon capture & 0.0019 &  Negligible \\
Neutron emission after muon capture & <0.001 & Negligible \\
Charged particle emission after muon capture & <0.001 & Negligible \\
Prompt beam induced electrons & <0.0038 & 0.002 \\
Radiative pion capture & 0.0028 & 0.001 \\
Delayed beam induced\tablefootnote{Assuming $R_\mathrm{ext}$ is $10^{-10}$ and $10^{-11}$ for Phase-I and Phase-II, respectively} & Negligible & 0.001 \\
Antiproton induced & 0.0012 & 0.30 \\
Cosmic ray induced & <0.01 & 0.29 \\
\hline
Total & <0.032 & 0.662 \\
\hline
\end{tabular}
\end{table}
The efficiencies have been obtained from a detailed simulation using the results of the tests on the prototypes.
The estimate of the number of events assumes 150 (230) days of data acquisition with 3.2 kW (56 kW) beam power in Phase-I (Phase-II) and $4.7\times 10^{-4}$ ($1.6\times 10^{-3}$) stopped muons per proton on target.
The single event sensitivity is defined as:
\begin{equation}
SES(\mu^-N \to e^-N) = \frac{1}{N_\mu \cdot f_{cap} \cdot f_{gnd} \cdot A_{sig}},
\end{equation}
where $N_\mu$ is the total number of stopped muons, $f_{cap}=0.61$ is the fraction of nuclear muon captures, $f_{gnd}=0.9$ is the model dependent fraction of muons to be the ground state.
Given the signal efficiencies reported in Table~\ref{tab:phys}, the expected S.E.S. for Phase-I (Phase-II) is $3 \times 10^{-15}$ ($1.4 \times 10^{-17}$), which is 100 (10,000) times better than the one achieved by SINDRUM II.

\section{Status}
The tests of the 8~GeV BSX proton beam have shown that the extinction factor is enough to keep the out of pulse background at a negligible level \cite{extinction}.
The C-line construction was recently completed up to the COMET beam area and is ready to provide the beam.
The coils for the proton capture solenoid are all winded and the cryostat cold bore part has been constructed \cite{magnet}.
The first 90 degrees of the muon transport solenoid, that is the full TS for Phase-I, has been constructed and tested up to 1.5~T.
The coil winding for the detector solenoid has been completed and the cryostat will be made soon this year.
The CDC has already been constructed, and its performance has been evaluated with cosmic rays.
It meets the performance requirement, a spatial resolution of 170~$\mu$m \cite{cdc}.
We have almost completed the R\&D for the CTH and the design has been fixed.
The support structure has been partially purchased and the detector construction is underway.
A trigger system for COMET Phase-I is being prepared and we have recently performed a successful test of a small trigger electronics chain.

The first station of the straw tube tracker has been completed \cite{strawConst} and commissioned in Phase-alpha beam measurement very recently.
The LYSO crystals are almost fully purchased, and the ECAL support structure has been manufactured.

The CRV design is almost finalised and the first module has recently been been produced.

For Phase-II, the thinner straw tube will be used with 12~$\mu$m with a twice smaller diameter.
The test production of thin straw tubes was done in success, and they are under the long term stability test \cite{thinstraw}.

In February and March 2023, the proton beam has been successfully delivered to the COMET beam area as COMET Phase-alpha. 
In this measurement, pions produced in a thin graphite target have been transported through the MTS and the muons coming from their decays have been successfully observed by a set of dedicated detectors.
Preliminary results can be found in \cite{kou_LP} while the final ones will be published soon.

\section{Summary and Prospects}

The COMET experiment aims to search for the $\mu$-$e$ conversion with 100 and 10,000 better sensitivities than the present upper limit.
The COMET beam line has been completed in 2022.
More recently the proton beam has successfully beam delivered to COMET Phase-alpha and the first muons have been observed. 
Phase-I preparation is well on track and first physics measurements are expected for 2024-2025.


\acknowledgments

This was written on behalf of, and the author acknowledges strong support from, the COMET collaboration.
We acknowledge support from JSPS, Japan; ARC, Australia; Belarus; NSFC, China; IHEP, China; IN2P3-CNRS, France; CC-IN2P3, France; SRNSF, Georgia; DFG, Germany; JINR; IBS, Korea; RFBR, Russia; STFC, United Kingdom; and the Royal Society, United Kingdom.
The views expressed herein are those of the author and are not necessarily those of the Australian Government or Australian Research Council.




\end{document}